\documentclass[doc,apacite]{apa6}\usepackage[]{graphicx}\usepackage[]{xcolor}
\makeatletter
\def\maxwidth{ %
  \ifdim\Gin@nat@width>\linewidth
    \linewidth
  \else
    \Gin@nat@width
  \fi
}
\makeatother

\definecolor{fgcolor}{rgb}{0.345, 0.345, 0.345}

\usepackage{framed}
\makeatletter
 {\par\unskip\endMakeFramed%
 \at@end@of@kframe}
\makeatother

\definecolor{shadecolor}{rgb}{.97, .97, .97}
\definecolor{messagecolor}{rgb}{0, 0, 0}
\definecolor{warningcolor}{rgb}{1, 0, 1}
\definecolor{errorcolor}{rgb}{1, 0, 0}
\newenvironment{knitrout}{}{} 

\usepackage{alltt}
\usepackage{amsmath,alltt,bm,url,subcaption}
\usepackage[english]{babel}

\title{Identifying good forecasters via adaptive cognitive tests}
\sixauthors{Edgar C.\ Merkle}{Nikolay Petrov}{Sophie Ma Zhu}{Ezra Karger}{Philip E.\ Tetlock}{Mark Himmelstein}
\sixaffiliations{University of Missouri}{Forecasting Research Institute\\ University of Cambridge}{Forecasting Research Institute\\ University of British Columbia}{Forecasting Research Institute\\ Federal Reserve Bank of Chicago}{Forecasting Research Institute\\ University of Pennsylvania}{Georgia Institute of Technology\\ Forecasting Research Institute}

\abstract{Assessing forecasting performance is a time intensive activity, often requiring months or years before we know whether or not the reported forecasts were accurate. Cognitive tests can be quickly administered and are predictive of forecasting performance, but it is unclear which and how many tests are optimal. In this study, we develop adaptive cognitive tests that optimize the selection and efficiency of cognitive tests to assess forecasters of different skill levels. The tests are based on item response models and the adaptive testing procedures commonly used in educational testing. We show how the procedures can select highly informative cognitive tests from a larger battery of tests, thereby reducing the time taken to administer the tests. We use a second, independent dataset to show that the selected tests yield scores that are highly related to out-of-sample forecasting performance. The approach enables real-time, adaptive testing, providing immediate insights into forecasting talent in practical contexts.}

\authornote{Correspondence to Edgar Merkle, 219 McAlester Hall, Columbia, MO 65211. Email: merklee@missouri.edu. Study materials are available at \url{https://github.com/forecastingresearch/fpt}. Scripts for reproducing the results of this paper are available at \url{https://github.com/ecmerkle/cog_adapt}.}
\shorttitle{Adaptive cognitive tests}

\let\proglang=\textsf

\IfFileExists{upquote.sty}{\usepackage{upquote}}{}
\begin{document}
\maketitle

\section{Introduction}
Researchers and practitioners have long relied on human forecasters to provide information about the likelihood of future events. These forecasts are especially useful for non-repeatable events in, say, geopolitics or popular culture, where there is little existing data that can be used to build predictive models of such events \cite<e.g.,>{atares17,walbud83}. Alternatively, when data do exist to build predictive models, human forecasts can be just as accurate as the models \cite{bejhyb2023, zelata2021}. In many cases, human forecasts can be less costly than the expertise and technology required for predictive modeling, often with little sacrifice in accuracy. 

When recruiting human forecasters, we immediately take interest in individual differences between the forecasters. Previous research \cite{melsto15,melung14} has demonstrated the existence of highly skilled forecasters who provide accurate forecasts across domains, which \citeA{tet16} termed {\em superforecasters}. We want to focus our resources on these exceptionally accurate forecasters, so that we can obtain accurate and timely aggregate forecasts \cite<e.g.,>{himtime2023}. But it often takes months or years for forecasting questions to resolve, which makes it difficult to assess individual differences in forecaster accuracy. One solution to this dilemma involves comparison of individual forecasters to their peers \cite<e.g.,>{budche15,chebud16,himwis2023,witata17}, prior to knowing the outcomes of the forecasting questions. We consider a second solution in this paper, which involves the use of cognitive tests to predict forecasting accuracy. 

Cognitive tests are advantageous for our purposes because they can be quickly scored. Past research has shown that the tests are correlated with forecasting accuracy \cite{atahim22,himata21, melpsyc15}, making it possible to quickly obtain information about which forecasters are likely to be accurate. But there are many potential cognitive tests that one {\em could} use, without much information about which tests one {\em should} use. Further, specific cognitive tests may serve specific purposes for predicting forecasting accuracy. For example, certain cognitive tests may help identify forecasters who lack any proficiency, while others may help distinguish top-tier forecasters. If we administer a large set of cognitive tests to all forecasters, then we waste resources when each test is only relevant for some of the forecasters. Further, if we want information about good forecasters as opposed to bad forecasters, then the good forecasters should complete more cognitive tests than the bad forecasters. We need ways to clarify which cognitive tests are good for which individuals, and to tailor administered tests to each individual's performance.

This paper introduces adaptive cognitive testing procedures, using psychometric models to tailor cognitive tests to individual forecasters and to predict their forecasting skill levels. The idea of selecting tests and tailoring them to individual forecasters is related to adaptive testing procedures that have been developed in the context of item response modeling \cite<e.g.,>{draols99,magrai12,meiner99,vangla00,wai00}. Many readers will have experienced adaptive tests that measure readiness for university studies. We extend these traditional adaptive testing procedures to the cognitive testing domain, and we then apply the procedures to predict forecasting accuracy across two independent sets of forecasters. We show how our adaptive cognitive tests can yield accurate assessments of forecasters' skill levels in reduced time, and we also obtain results about which cognitive tests are generally better or worse for predicting forecasting accuracy across all skill levels.

The paper is organized as follows. We first give a conceptual overview of traditional adaptive testing methods. We then show how the methods can be extended to cognitive testing. Next, we apply the model to two independent datasets, identifying cognitive tests that effectively predict different levels of forecasting skill. Next, we develop a fully adaptive testing procedure that tailors the sequence of cognitive tests to forecasters' performance, and we show that the procedure can reduce the time spent on cognitive testing. Finally, we discuss limitations and potential future developments.

\section{Adaptive Testing Overview}
In this section, we give a brief conceptual overview of adaptive testing. We then discuss the technical details in the following section. The models that we consider are related to some that have been previously applied to forecasting by \citeA{merste16,merste17} and \citeA{bobud17}, though those authors did not consider adaptive tests.

The idea of adaptive testing is to choose a question to present to a person during a test, based on known question characteristics and based on the person's previous responses during the test. Standardized tests like the GRE or GMAT are examples of adaptive testing. Adaptive testing is based on the notion that different questions can provide different amounts of information about different individuals. For example, if a person has already answered many questions incorrectly, then little is gained from presenting that person with a difficult question. We already know that the person will probably answer incorrectly. Similarly, if a person has already answered many questions correctly, we do not learn much by presenting that person with an easy question. By adaptively presenting questions that are uniquely suited to each individual, we can potentially obtain a precise estimate of their proficiency using shorter tests.

Adaptive testing works by first estimating question properties during an ``item calibration'' step. Calibration involves determining properties of each test question (e.g., the question's difficulty), with these properties being used in later steps of the adaptive testing procedure. In this first step, no adaptive testing is used. We simply present the questions to a large sample of representative people, fit item response models to their data, and treat the resulting parameter estimates as known.
Once we have the parameter estimates, we can use those to compute question {\em information functions}. These functions summarize the amount of information that each question provides about participants of different proficiencies. 
They formalize the ideas presented in the previous paragraph, where a difficult question provides more information about a person of high proficiency as compared to a person of low proficiency, and vice versa. Once we have the question information functions, they can be used to adaptively select questions. That is, at each point in a test, we can compute an estimate of a person's proficiency. With that estimate, we can use question information functions to determine which of the remaining questions will provide the most information about the person.

Question information functions are related to the Fisher information matrix that is widely used to compute standard errors of model parameters estimated via maximum likelihood. The intuition is that a high amount of question information implies a smaller standard error, which in turn implies a more precise estimate of a person's proficiency.
From a technical standpoint, question information functions involve second derivatives of an item response model's log-likelihood, where the derivatives are taken with respect to each person's proficiency parameter (a parameter that is often denoted $\theta$ in item response models). These derivatives have analytic forms for many item response models \cite<e.g.,>{wangra22}, which makes them suited for the adaptive testing purposes considered here.

\section{Application to Forecasting}
The discussion so far has focused on individual questions administered in a single test. These individual questions are often scored as ``correct'' or ``incorrect.'' By contrast, for many cognitive tests, only an overall score is recorded. In this case, we can treat each cognitive test as an individual question, with information functions indicating which tests are best suited to high- and low-proficiency forecasters.

We immediately encounter the problem of choosing a suitable item response model. For many cognitive tests, each person's score is an integer that could potentially be treated as continuous. The simplest thing to do would be to treat the scores as continuous and fit a Gaussian model, where the model has question and person parameters similar to traditional item response models. But this ignores the fact that most cognitive tests have a predefined lower and upper bound.

To better characterize our bounded test scores, we employ a beta-distributed item response model \cite{noedau07}. The beta distribution is particularly suited for bounded cognitive test scores, where a defined score range typically exists. The model's technical details remain similar to traditional item response models, except the model likelihood involves the beta distribution instead of the binomial distribution. We specifically use a variation of the \citeA{noedau07} model that was recently described by \citeA{molcur22}. For general information about modeling data via the beta distribution, see \citeA{smiver06}.

\subsection{Model Definition}
Let $y_{ij}$ be the score of participant $i$ on cognitive test $j$, where the scores have been scaled to lie in $[0,1]$. Let $z_{ij}$ be a three-category ordinal variable indicating whether $y_{ij}$ is at a boundary or between the boundaries (1 if $y_{ij}=0$; 2 if $y_{ij}$ is between 0 and 1; 3 if $y_{ij}=1$). Our model has two parts: one part that predicts $z_{ij}$, and one part that predicts values of $y_{ij}$ between 0 and 1. As further described below, some parameters are shared between the two parts of the model.

\paragraph{Submodel for $z_{ij}$}
Following \citeA{molcur22}, we use a graded response model to predict the chance that a person's score is at the lower boundary, the upper boundary, or in between. This part of the model can be written as
\begin{equation*}
P(z_{ij} \leq c | \theta_i) = \text{logit}^{-1}(\gamma_{jc} - \alpha_j \theta_i)\ \ c=1,2,
\end{equation*}
where $\gamma_{j1} < \gamma_{j2}$. We obtain probabilities of assuming individual categories by subtracting successive terms and making use of the fact that the three probabilities must sum to 1:
\begin{align*}
  P(z_{ij} = 1 | \theta_i) &= \text{logit}^{-1}(\gamma_{j1} - \alpha_j \theta_i) \\
  P(z_{ij} = 2 | \theta_i) &= \text{logit}^{-1}(\gamma_{j2} - \alpha_j \theta_i) - \text{logit}^{-1}(\gamma_{j1} - \alpha_j \theta_i) \\
  P(z_{ij} = 3 | \theta_i) &= 1 - \text{logit}^{-1}(\gamma_{j2} - \alpha_j \theta_i).
\end{align*}
The $\alpha_j$ parameter is commonly called the discrimination parameter, while the $\gamma_j$ parameters help account for the base rates of boundary and non-boundary responses. Further, $\text{logit}^{-1}(x) = (1 + \exp(-x))^{-1}$ is the inverse logit link function, which converts an unbounded prediction to a prediction that lies in $(0,1)$.

\paragraph{Submodel for Non-Boundary Responses}
When $z_{ij} = 2$, the response $y_{ij}$ could lie anywhere between 0 and 1. We use a beta item response model here, where the $\alpha_j$ and $\theta_i$ parameters are shared with the graded response submodel described above. The beta item response model assumes independence of test scores given $\theta_i$:
\begin{equation}
   y_{ij} \mid \theta_i \sim \text{Beta}(a_{ij}, b_{ij}),
\end{equation}
where the beta distribution follows the traditional parameterization, with two positive shape parameters $a$ and $b$. Under this parameterization, our model assumes the following, unintuitive form:
\begin{align*}
    a_{ij} &= \exp \left ( \frac{\beta_j + \alpha_j \theta_i + \omega_j}{2} \right ) \\
    b_{ij} &= \exp \left ( \frac{-(\beta_j + \alpha_j \theta_i) + \omega_j}{2} \right ).
\end{align*}
These expressions are relatively simple, and they make it clear that there are three parameters per test (those with a $j$ subscript) and one parameter per person ($i$ subscript). But it is not so easy to recognize this as an item response model.

For the purposes of understanding the model, we write the mean of the beta distribution as $\mu_{ij} = a_{ij}/(a_{ij} + b_{ij})$. The expressions above then lead us to a traditional, two-parameter item response function on $\mu_{ij}$ \cite<see>{noedau07}:
\begin{equation}
    \mu_{ij} = \text{logit}^{-1}(\beta_j + \alpha_j \theta_i),
\end{equation}
where $\beta_j$ is often called the ``easiness'' parameter, and $\alpha_j$ is the discrimination parameter from the graded response submodel. 

Because the beta distribution is bounded on both sides, the variance of the beta distribution is partly influenced by $\mu_{ij}$. To account for response heterogeneity across cognitive tests, our third test parameter $\omega_j$ helps predict the precision (inverse variance) of the beta distribution. With this parameter, the variance of the beta distribution can be written as
\begin{equation*}
    \text{Var}(y_{ij} | \theta_i) = \frac{\mu_{ij} (1 - \mu_{ij})}{1 + 2\exp(\frac{\omega_j}{2})\cosh(\frac{\beta_j + \alpha_j \theta_i}{2})},
\end{equation*}
where the variance becomes larger when $\mu_{ij}$ is near 0.5 and when $\omega_j$ is negative.

\subsection{Information Functions}
Combining the results from the previous sections, we can write the overall model as
\begin{equation}
   \label{eq:lik}
  p(y_{ij} | \theta_i) = \begin{cases}
    \text{logit}^{-1}(\gamma_{j1} - \alpha_j \theta_i)& y_{ij} = 0 \\
    \left ( \text{logit}^{-1}(\gamma_{j2} - \alpha_j \theta_i) - \text{logit}^{-1}(\gamma_{j1} + \alpha_j \theta_i) \right ) \times \text{Beta}(a_{ij}, b_{ij})& 0 < y_{ij} < 1 \\
    1 - \text{logit}^{-1}(\gamma_{j2} - \alpha_j \theta_i)& y_{ij} = 1.
    \end{cases}
\end{equation}
The information function for each cognitive test, which can be used for adaptive testing purposes, is obtained by taking second derivatives of the resulting model likelihood with respect to $\theta_i$. \citeA{molcur22} shows that the information function is
\begin{equation*}
  \begin{split}
  I_j(\theta) = \alpha_j^2 & \left ( Q_0^2 P_0 + P_1^2 Q_1 - (P_1 - P_0) \left ( \frac{P_1 Q_1 - P_0 Q_0}{P_1 - P_0} \right )^2 - \right. \\
  & \ \left. (P_1 - P_0) \left [ \Omega(a_{ij} + b_{ij}) \left ( \frac{a_{ij} - b_{ij}}{2} \right )^2 - \Omega(a_{ij})\left ( \frac{a_{ij}}{2} \right )^2 - \Omega(b_{ij}) \left ( \frac{b_{ij}}{2} \right )^2 \right ] \right ),
  \end{split}
\end{equation*}
where
\begin{align*}
  P_0 &= \text{logit}^{-1}(\gamma_{j1} - \alpha_j \theta_i) \\
  Q_0 &= 1 - \text{logit}^{-1}(\gamma_{j1} - \alpha_j \theta_i) \\
  P_1 &= \text{logit}^{-1}(\gamma_{j2} - \alpha_j \theta_i) \\
  Q_1 &= 1 - \text{logit}^{-1}(\gamma_{j2} - \alpha_j \theta_i),
\end{align*}
and $\Omega()$ is the trigamma function, which is a well-known mathematical series that can be closely approximated by most statistical software.

\subsection{Model Discussion}
In this subsection, we clarify and justify some specific modeling choices that we made. These choices involve the handling of boundary test scores, and the assumption that proficiency is unidimensional. We address each choice separately below.

First, our model follows the work of \citeA{molcur22} in that we explicitly model boundary responses. In traditional models involving the beta distribution, exact values of 0 and 1 are impossible, and it is customary to change 0s to 0.0001 or some other small value, and to change 1s to 0.9999 or some other large value. The cognitive tests that we employ in this paper do not always have hard boundary values, and we instead use empirical boundaries, i.e., the smallest and largest observed scores on each test. The use of empirical boundaries is imprecise because there remains the possibility that future respondents receive scores that are more extreme than the empirical boundaries. Such extreme scores would be censored by the empirical boundaries that were already set, which makes it more difficult to assess respondents who are exceptionally skilled or exceptionally unskilled. While our current treatment of boundaries works reasonably well in the applications, we further consider this issue in the General Discussion.

Second, our model assumes that each person's proficiency is unidimensional. In other words, the model assumes that a single number fully describes each person's proficiency across all the cognitive tests. This assumption is surely incorrect, because there are many aspects of a person's proficiency that may lead them to do better or worse on specific tests. But we make this assumption for two reasons. First, our overall goal is to use the cognitive tests to gauge future forecasting accuracy, so that it is beneficial to have a one-number summary of each person's proficiency. In other words, we have the model reduce all the cognitive test scores into the best one-number summary possible. Second, additional dimensions of proficiency can lead to noisier estimates for each dimension, which may hamper our ability to predict future forecasting accuracy. We also return to this issue in the General Discussion.

\section{Study 1: Test Information and Correlates with Forecasting}
The datasets that we use to develop adaptive tests were collected as part of a large project on forecasting proficiency assessment \cite{himzhu24}. Our first study involves data from an exploratory pilot experiment where 170 participants completed a variety of cognitive tests and forecasting tasks, details of which appear below. Our second study, discussed later, involved over 1,000 participants and was similar to the pilot. The two studies allow us to compare the stability of model estimates over time and to examine out-of-sample model predictions. But for our first study, we examine the model's fit to the cognitive test scores, as well as the information functions for each cognitive test.

\subsection{Method}
Study 1 included a screening survey, followed by three weekly surveys, with a total of 170 participants being recruited on Facebook and responded to the surveys. One of the three surveys (after screening) involved forecasting tasks, while the other two surveys involved the cognitive tests that are of interest here. The cognitive tests were all intended to be simple measures of cognitive reasoning and/or numeracy. Table~\ref{tab:ctests} shows the tests that were used, along with key references for the tests, with further information about the rationale for these tests being found in \citeA{himzhu24} as well as the OSF repository at \url{https://osf.io/q2e94/}. The Adult Decision Making Competence test has three subscales, the Shipley test has two subscales, the Impossible Question test has three separate scores (one for the impossible questions and two for the ``possible,'' regular questions), and the Bayesian Updating and Denominator Neglect tasks each have two versions. 

The forecasting questions were selected from a database of questions used in previous forecasting tournaments \cite{zouxia22}, with each participant forecasting 15 questions. The questions involved a range of topics including economics, politics, and popular entertainment, all of which could be understood by laypeople. All of the questions involved time series, so that participants could always predict the value of the time series at a future date, and so that the questions could be reused for different resolution dates. The specific questions used can be found in the project materials.

Forecasts were elicited via quantile judgments, with forecast accuracy being measured via an {\em S-score} \cite{che22,grulic17,joswin09}. This is a strictly proper scoring rule for quantile forecasts that is based on the absolute distance between each quantile estimate and the realized outcome, with larger penalties being imposed when the outcome is extreme relative to the quantile estimates. Forecaster accuracy was summarized via an average S-score across questions. After fitting our model to the cognitive test scores, we examined the relationship between participants' performance on the cognitive tests and their average S-score.

\begin{table}
  \caption{Cognitive tests and primary references.} \label{tab:ctests}
  \begin{center}
    \begin{tabular}{ll}\hline
      Test Name & References \\\hline
      Adult Decision Making Competence & \citeA{bru07} \\
      Bayesian Updating & \citeA{phi66} \\
      Berlin Numeracy & \citeA{cokgal12} \\
      Cognitive Reflection & \citeA{top13} \\
      Coherence Forecasting & \citeA{hobud24} \\
      Denominator Neglect & \citeA{kir92} \\
       & \citeA{mik15} \\
      Graph Literacy & \citeA{gal10} \\
      Impossible Question & \citeA{ben22} \\
      Leapfrog & \citeA{knox12} \\
      Raven Matrices & \citeA{mat10} \\
      Number Series & \citeA{die16} \\
       & \citeA{himata21} \\
      Shipley General Intelligence & \citeA{ship09} \\
      Time Series & \citeA{debvan21}\\
      & \citeA{reihar11} \\\hline
    \end{tabular}
  \end{center}
\end{table}

Our sample size of 170 participants is small for item response applications, an issue that we rectify in Study 2. About 4\% of test scores were missing, because participants sometimes started a session but did not finish it. We did not do anything sophisticated to model missingness; model estimation involves the traditional ``missing at random'' assumption, where we use all observed data and ignore missing data. We fit the model to the test scores via Markov chain Monte Carlo with Stan, saving 1,000 posterior samples from each of 3 chains after 1,000 warmup iterations. Model convergence was judged by the Rhat and effective sample size metrics \cite<e.g.,>{vehgel21}, neither of which indicated estimation problems.

In our model, we used mildly informative prior distributions that constrain parameters to plausible values of the parameter space. Our specific prior distributions were
\begin{align*}
  \beta_j &\sim \text{N}(0,2) \\
  \alpha_j &\sim \text{logN}(0,1) \\
  \omega_j &\sim \text{N}(0,10) \\
  \gamma_{j1} &\sim \text{N}(-2,1) \\
  \gamma_{j2} &\sim \text{N}(2,1),
\end{align*}
where normal distributions are parameterized via standard deviations and where $j=1,\ldots,20$. These priors are on the logit scale, which is similar to the z-score (or probit) scale. Accordingly, we generally expect model predictions to be between $-4$ and $+4$ (on the logit scale), with increases of 1 being fairly large. Thus, the above priors are not as informative as they may appear at first glance.

\subsection{Results}
The appendix includes pairwise correlations between all 20 of the cognitive tests (treating the test scores as continuous), which is informative about the degree to which each individual test is related to the others.
We focus here on our modeling results, which are divided into three subsections below. First, we examine overall fit of the model to the data. Next, we examine the information provided by each cognitive test, as estimated by our item response model. Finally, we use the estimated test information to select specific tests, and we examine how the selected tests are related to forecasting accuracy.

\begin{figure}
  \caption{Posterior predictive checks of fitted model. The black curve represents the observed distribution of scores on each test, and the light blue lines represent posterior predictions.}  \label{fig:zrep}
\begin{knitrout}\footnotesize
\definecolor{shadecolor}{rgb}{0.969, 0.969, 0.969}\color{fgcolor}

{\centering \includegraphics[width=\maxwidth]{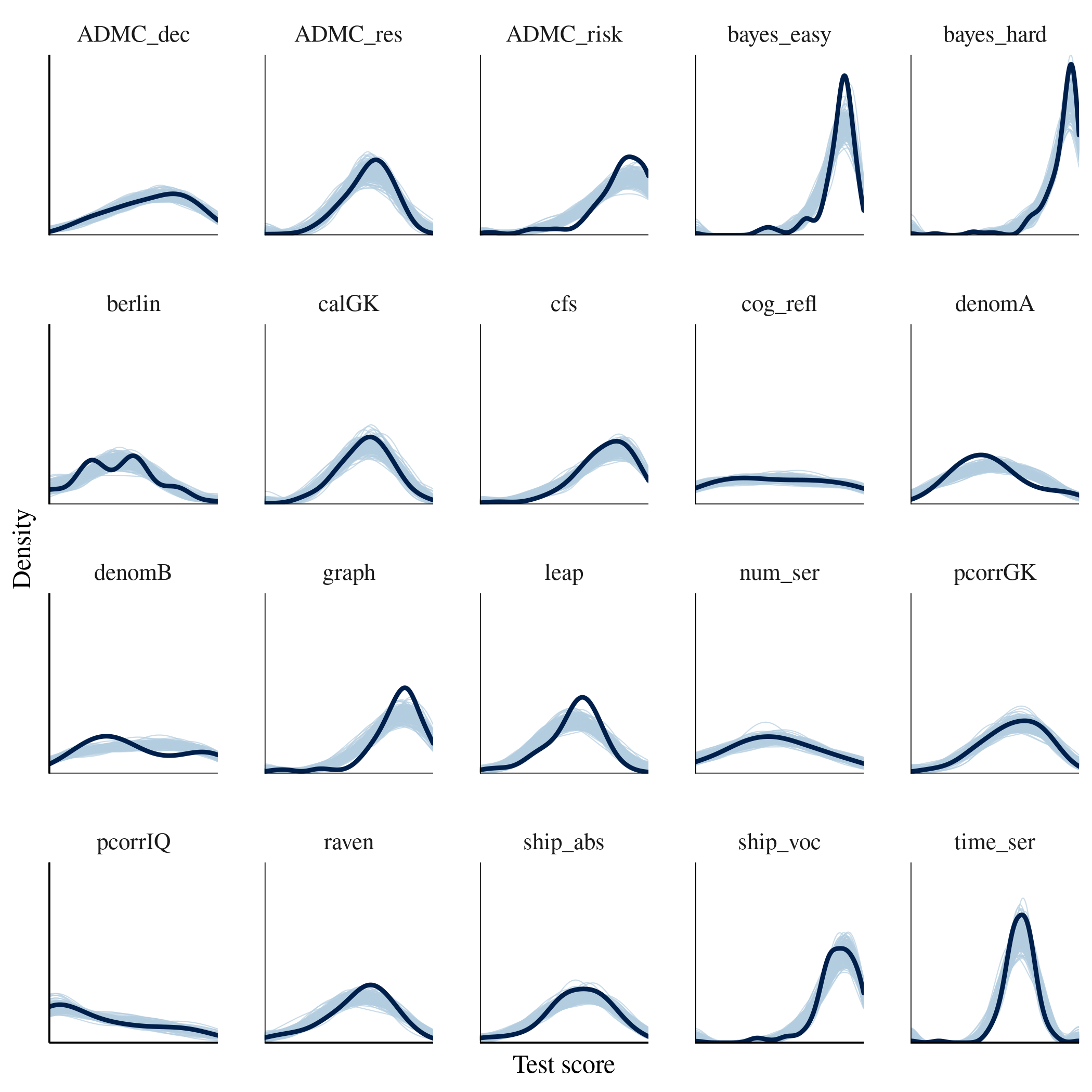} 

}

\end{knitrout}
\end{figure}

\paragraph{Model Fit}
Figure~\ref{fig:zrep} is a posterior predictive check to ensure that the model is describing the data reasonably well. In each panel of this figure, the black curve is a smoothed histogram that represents the observed distribution of scores on each cognitive test. The light blue curves represent the model's posterior predictive distribution of the scores, which collectively represent uncertainty in model predictions. The figure shows that the model accurately reproduces cognitive test scores, which supports the idea that the model can be used for adaptive testing.

\begin{figure}
  \caption{Forecasting S-scores vs cognitive test proficiency.}  \label{fig:fscores}
\begin{knitrout}\footnotesize
\definecolor{shadecolor}{rgb}{0.969, 0.969, 0.969}\color{fgcolor}

{\centering \includegraphics[width=\maxwidth]{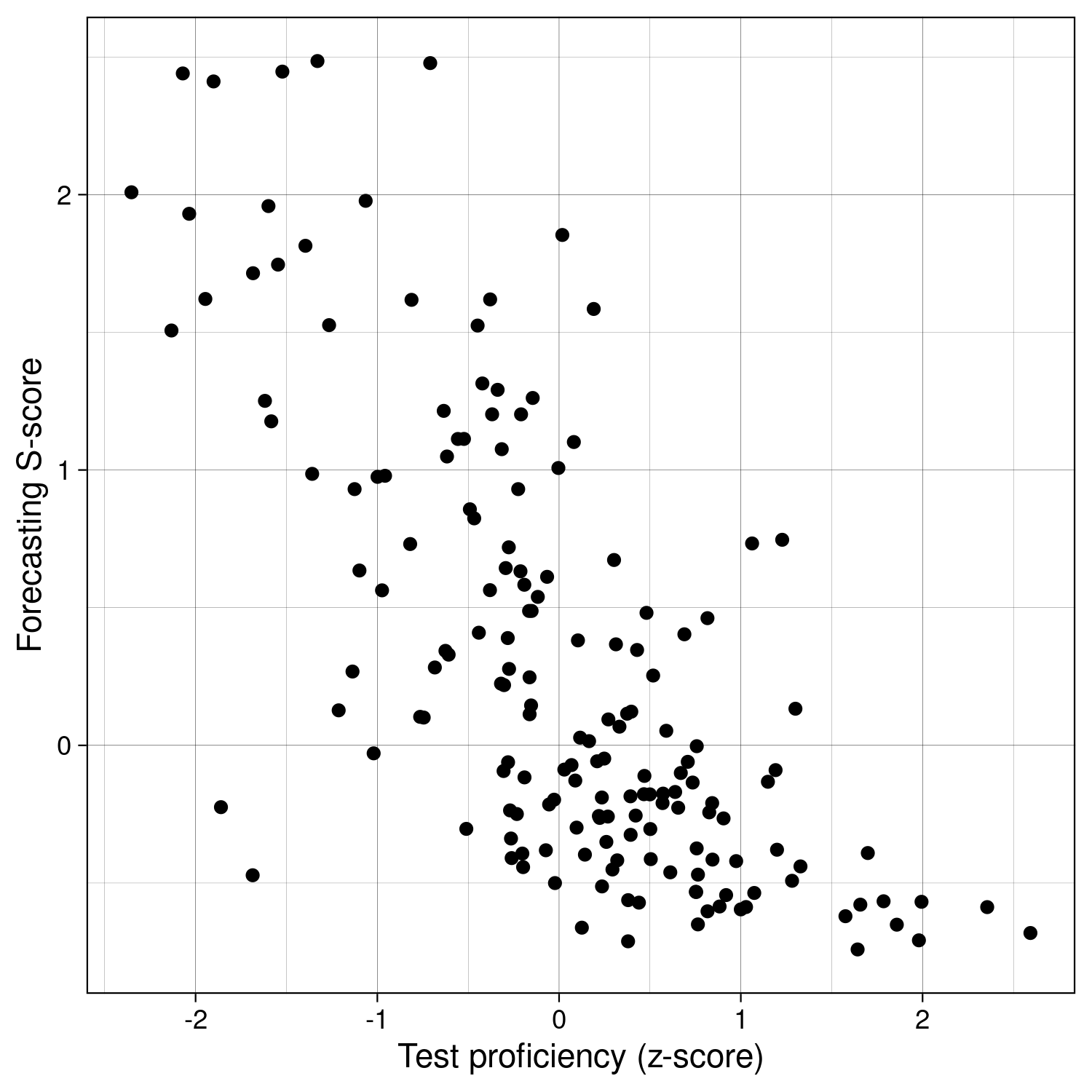} 

}

\end{knitrout}
\end{figure}

The model provides an estimate of each person's proficiency across the tests. If the model is to be useful, we should expect the test proficiency estimates to be related to each person's eventual forecasting performance. We obtained each person's average S-score, which represents a one-number summary of forecasting performance. We then examined the relationship between S-scores and the model's proficiency estimates from the cognitive tests. This relationship is important because it provides evidence that cognitive test proficiency is predictive of forecasting proficiency. The results are shown in Figure~\ref{fig:fscores}, where cognitive test proficiency is on the x-axis and forecasting S-scores are on the y-axis. We see a negative relationship, where lower S-scores are better and higher proficiency scores are better. The figure shows a correlation of $-0.71$, providing evidence that the model-based person estimates from the cognitive tests are strongly related to eventual forecasting performance. For comparison, we also computed an average standardized cognitive test score for each person, which does not involve any model estimates. The correlation between this average score and forecasting performance is $-0.7$. This helps build confidence that the model is providing a faithful representation of the data. More generally, these high correlations show that the cognitive test scores are predictive of eventual forecasting performance.

\begin{figure}
  \caption{Posterior predictive distributions of test information.}  \label{fig:postsampinf}
\begin{knitrout}\footnotesize
\definecolor{shadecolor}{rgb}{0.969, 0.969, 0.969}\color{fgcolor}

{\centering \includegraphics[width=\maxwidth]{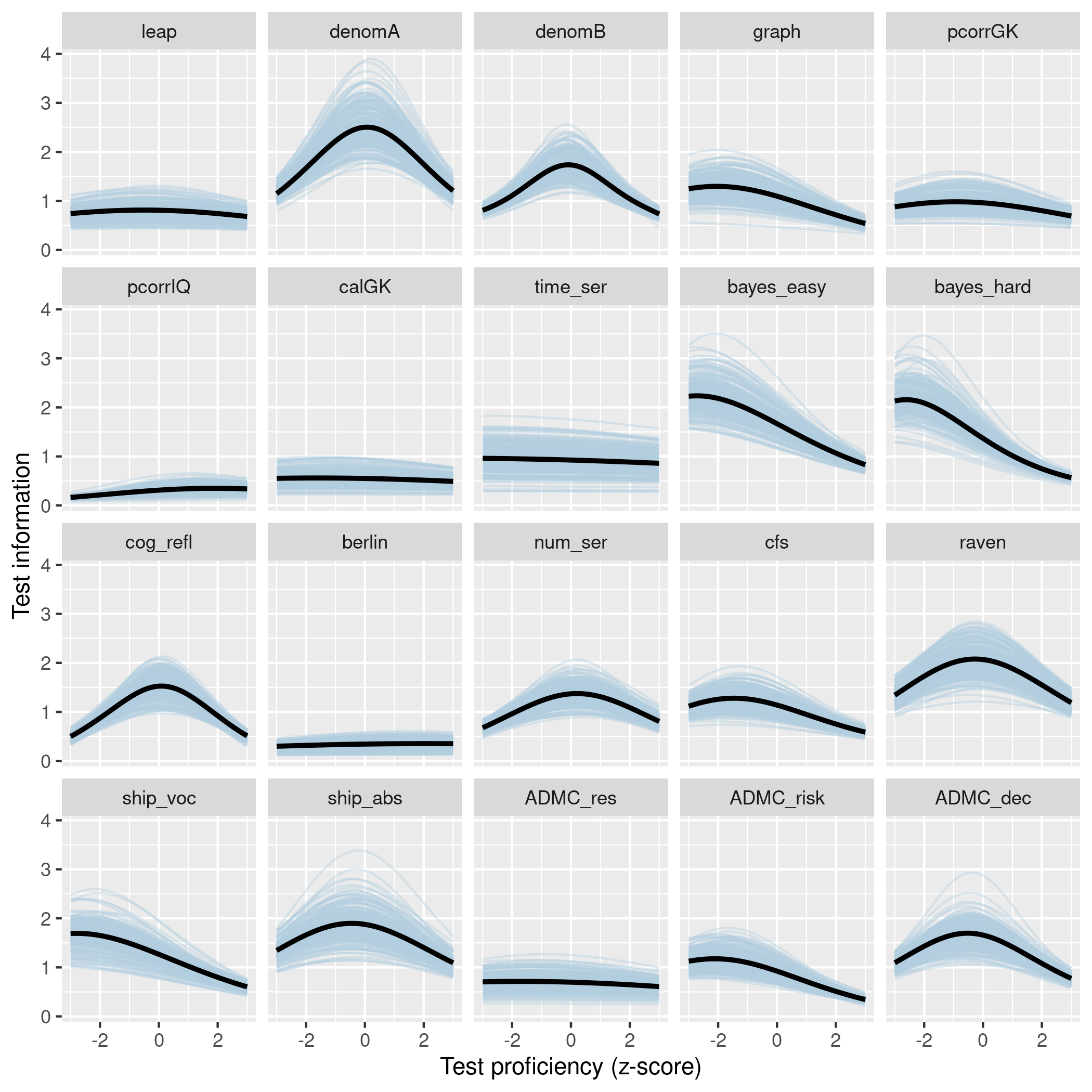} 

}

\end{knitrout}
\end{figure}

\paragraph{Test Information}
Given that the model is behaving reasonably, we turn to the information functions of each cognitive test.
Figure~\ref{fig:postsampinf} shows information separately for each cognitive test,
where cognitive test proficiency is on the x-axis and information is on the y-axis. Cognitive test proficiency is defined on the same scale as z-scores, where 0 represents a person at the 50th percentile and positive values represent people above the 50th percentile. The information values on the y-axis are bounded from below at 0, and larger values are better. The black curves show test information using posterior mean parameter estimates, while the light blue curves show posterior uncertainty.

Figure~\ref{fig:postsampinf} shows that the cognitive tests generally provide more information about people of lower proficiency, compared to people of higher proficiency. This is because the test information curves tend to be higher for proficiency values below 0. We see that Raven Matrices, Denominator Neglect, and Bayesian Updating are most informative, especially for people whose proficiency is near the 50th percentile. Those metrics also tend to be relatively informative about people of higher proficiency, though the magnitude of information decreases near the right side of each panel.
The figure also shows that uncertainty for uninformative scales (low and flat curves) tends to be low, compared to uncertainty for more informative scales (high and peaked curves). And while there is definitely uncertainty in cognitive test information, the figure shows that the uncertainty is not so high that we cannot distinguish between scales. For example, even after considering posterior uncertainty, we could still conclude that the Raven Matrices are more informative than the Berlin Numeracy test.

\begin{figure}
  \caption{Posterior predictive distributions of test information per minute.}  \label{fig:postsampmin}
\begin{knitrout}\footnotesize
\definecolor{shadecolor}{rgb}{0.969, 0.969, 0.969}\color{fgcolor}

{\centering \includegraphics[width=\maxwidth]{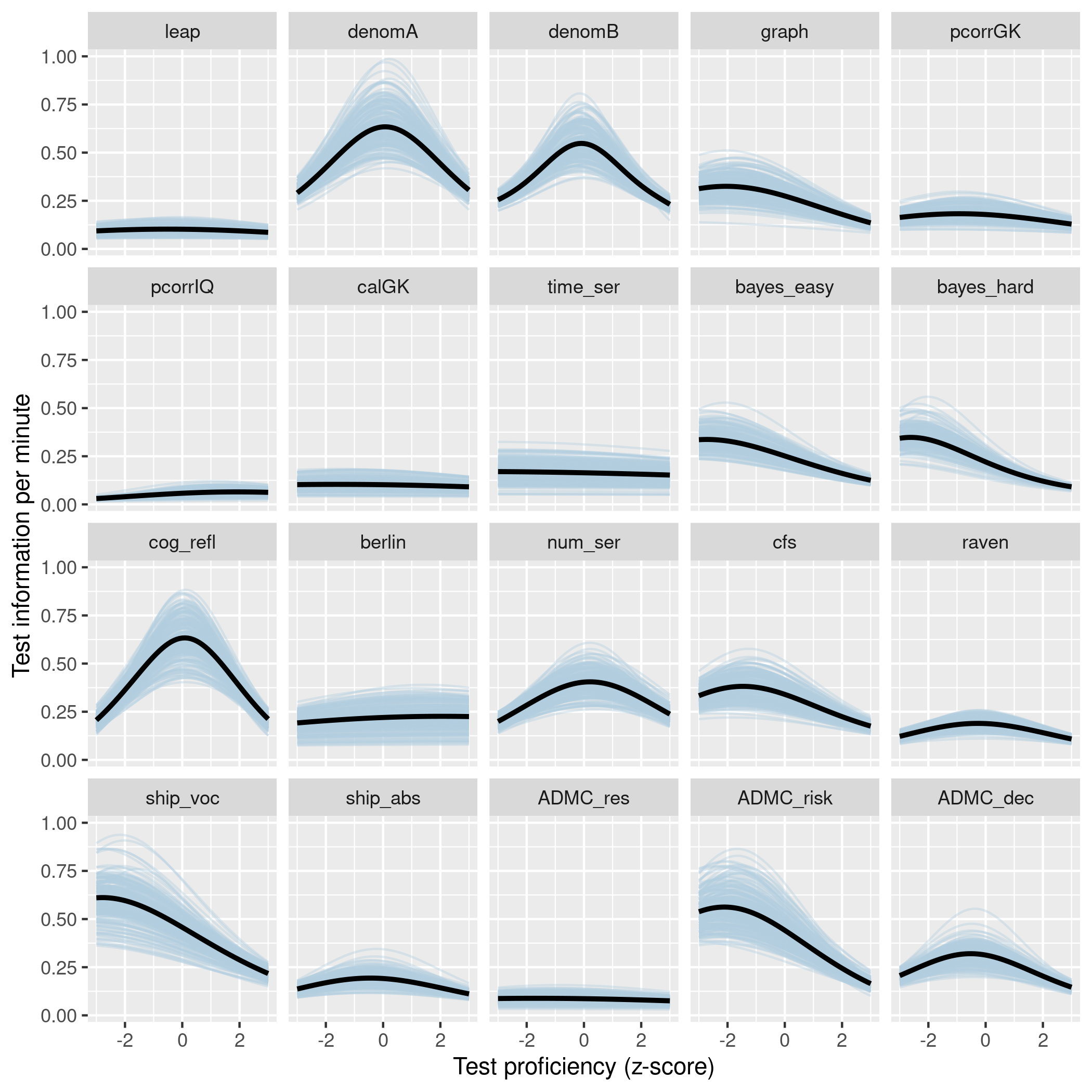} 

}

\end{knitrout}
\end{figure}

A problem associated with Figure~\ref{fig:postsampinf} is that it does not account for the amount of time that it takes to complete each test. If the tests with the most information each take a long time to complete, then we might still be better off administering a series of quick tests. To address this issue, we compute an ``information per minute'' metric that involves division of the Figure~\ref{fig:postsampinf} curves by the average time that each test takes to complete. Results are shown in Figure~\ref{fig:postsampmin}, which is arranged similarly to the previous figure. We now see that the Cognitive Reflection and Denominator Neglect tasks become especially informative because the curves are higher than those of other tests. The Cognitive Reflection and Denominator Neglect tests each average 2--3 minutes for completion, whereas other informative tests from Figure~\ref{fig:postsampinf} take a longer time. For example, Bayesian updating and Raven matrices average 6 minutes and 11 minutes, respectively.

\paragraph{Test Selection}
To select a subset of cognitive tests for administration, we focus on the ``question information per minute'' curves in Figure~\ref{fig:postsampmin}. If we want to administer tests that are most informative about an average forecaster, we look for the curves that are highest at a proficiency value of 0. These include Cognitive Reflection, Denominator Neglect, and Shipley Vocabulary. If we want to administer tests that are most informative about the top 10\% of forecasters, we look for the curves that are highest for (say) test proficiency near 1.3 (this is the z-score that cuts off the top 10\% of the distribution). These are generally the same tests, with the possible addition of Berlin Numeracy. There are also some tests that provide little information across the full range of proficiency (Leapfrog, Impossible Question, General Knowledge Calibration), suggesting that they can be removed from the battery of tests.

To begin to explore the utility of these ideas, we computed each participant's average standardized score on a subset of cognitive tests. Then we examined the relationship between these scores and their eventual forecasting S-scores. We expected that, if we maintained the most informative cognitive tests and discarded the uninformative tests, we could maintain a strong relationship with forecasting performance while reducing the amount of time that it takes to administer cognitive tests. Based on the results shown in Figure~\ref{fig:postsampmin}, we chose Cognitive Reflection, Denominator Neglect (A and B), Berlin Numeracy, Shipley Vocabulary, and Number Series as our subset of tests that provide the most information per minute. The time for completing this subset of tests is around 17 minutes (computed by summing median completion times for each individual test), whereas the time for completing all tests is around 103 minutes. 

\begin{figure}
  \caption{Relationships between forecasting S-scores, average cognitive test scores (scaled between 0 and 1), and average of highly informative cognitive test scores (scaled between 0 and 1).}  \label{fig:scatpl}
\begin{knitrout}\footnotesize
\definecolor{shadecolor}{rgb}{0.969, 0.969, 0.969}\color{fgcolor}

{\centering \includegraphics[width=\maxwidth]{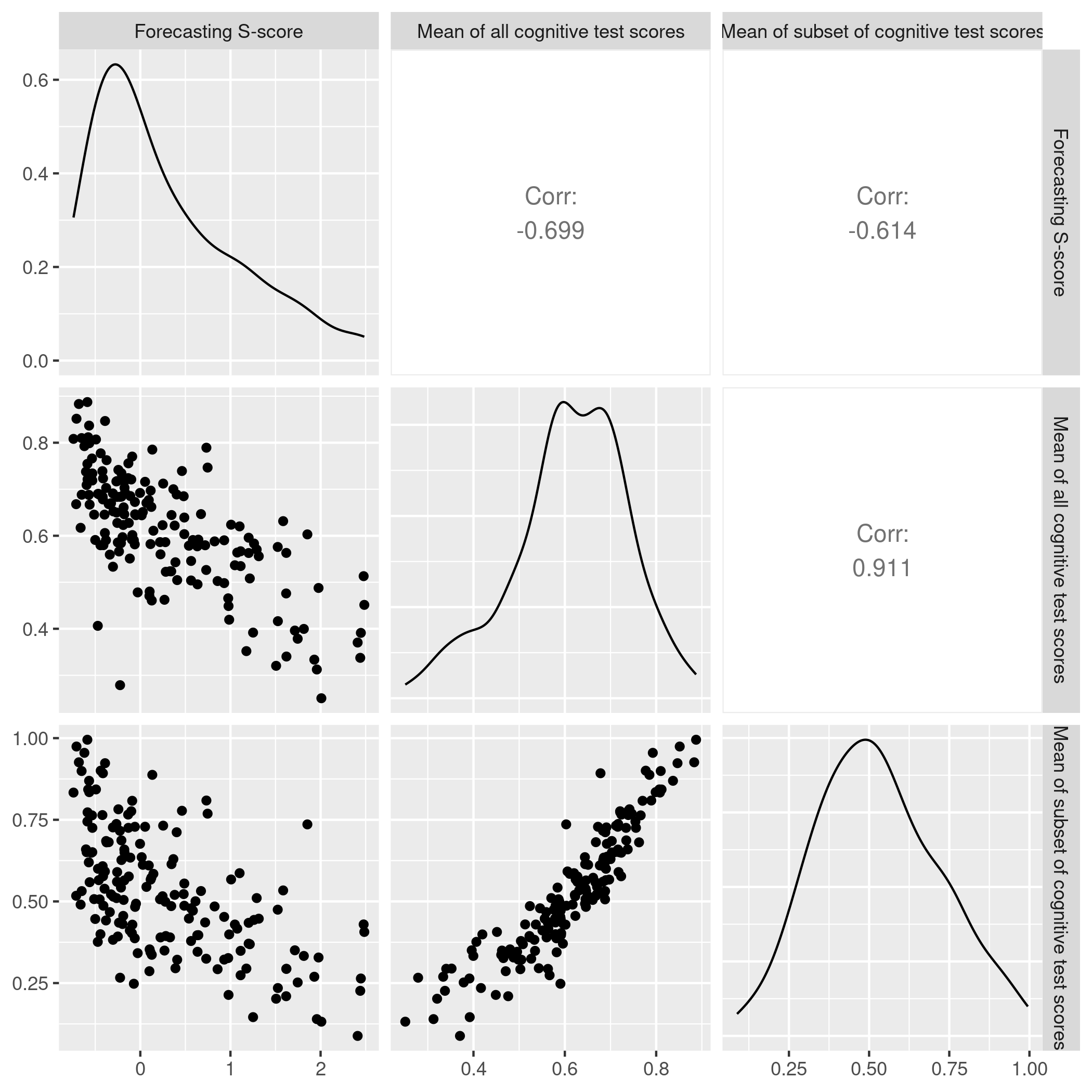} 

}

\end{knitrout}
\end{figure}

Figure~\ref{fig:scatpl} shows the relationships between forecasting accuracy (S-score), average standardized score across all cognitive tests, and average standardized score across the informative subset of cognitive tests. The figure shows that, when we use the selected subset of tests, the correlation with forecasting accuracy is reduced but remains strong overall: it goes from $-0.7$ when using all cognitive tests to $-0.61$ when using the subset. If we add Shipley Abstraction and Raven Matrices to our test subset, the correlation with forecasting accuracy becomes $-.67$ but the median time for taking the augmented subset of tests is now 38 minutes. This illustrates a tradeoff between cognitive test length and predictive ability, which might lead us to consider the desired length of time that participants should spend on cognitive testing.

\subsection{Summary}
Our item response model provided a reasonable fit to the Study 1 data, provided evidence of tests that were more and less informative about cognitive proficiency, and yielded scores that were highly predictive of later forecasting performance. But this was a small dataset compared to most item response applications. In the next section, we reproduce these results using a larger dataset, and we also compare model estimates across the two datasets.

\section{Study 2: Model Stability and Out-of-Sample Predictions}
Study 2 was a larger experiment that was collected about eight months after Study 1. This second study allows us to examine the stability of the model estimates across datasets, and it also allows us to score new participants using the test parameter estimates from Study 1. The analyses provide further evidence of the model's utility for the purposes of adaptive testing, where cognitive tests are automatically selected based on a participant's performance on previous tests.

\subsection{Method}
The data collection was similar to that of Study 1, with the involvement of more participants and more forecasting questions. The 1194 participants in this study forecasted 36 time series questions, many of which overlapped with those used in Study 1 (see the Supplemental Material; some question stems were the same in Study 1 and Study 2, but the time frames differed). Participants completed the same cognitive tasks that were used in Study 1, and our model priors and estimation details remained the same as Study 1. Conventional Bayesian diagnostics again indicated that the model converged. 

Now that we have two independent datasets, multiple options are available for scoring participants on the cognitive tests. As was done in Study 1, we can fit the item response model to our new dataset and use the estimated $\theta_i$ parameters as a measure of each participant's overall performance. Alternatively, we can use the existing question parameter estimates from Study 1 to score participants from Study 2. This scoring option is advantageous because it does not require us to fit a new model to the Study 2 data, so forecasters can be scored in real time as they complete cognitive tests. To do the scoring, we take the log-likelihood function implied by Equation~\eqref{eq:lik} and fix the question parameters to their posterior means from Study 1. We then numerically optimize the log-likelihood with respect to $\theta_i$, separately for each $i$ (where $i$ indexes Study 2 participants). This unidimensional optimization is much faster than fitting a full model to the new dataset.

\subsection{Results}
\begin{figure}
  \caption{Question parameter estimates for Study 1 versus Study 2. Error bars represent one posterior standard deviation around the posterior mean.} \label{fig:parcomp}
\begin{knitrout}\footnotesize
\definecolor{shadecolor}{rgb}{0.969, 0.969, 0.969}\color{fgcolor}

{\centering \includegraphics[width=\maxwidth]{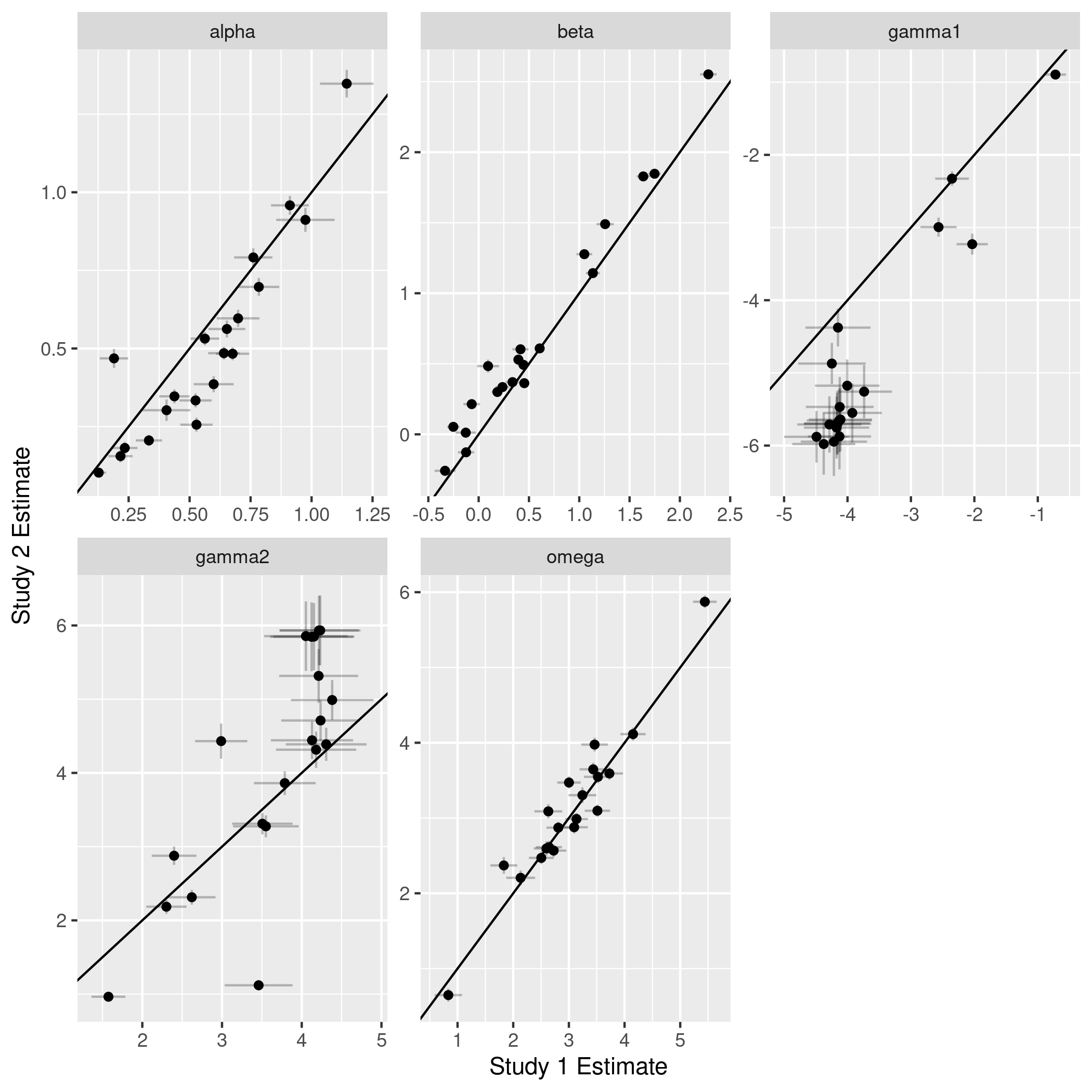} 

}

\end{knitrout}
\end{figure}

We begin by comparing the question parameter estimates across the two studies. Figure~\ref{fig:parcomp} provides this comparison, where each panel is a question parameter, Study 1 estimates appear on the x-axis, and Study 2 estimates appear on the y-axis. Each point represents one of the 20 cognitive tests that were administered in both Study 1 and Study 2, with the horizontal error bars being larger because there were fewer participants in Study 1 than in Study 2.

Examining the figure, we see that the easiness and discrimination parameters (alpha and beta) generally agree across studies, with most points falling near the diagonal identity line. The same is true of the omega parameters, which represent residual variability in the beta distribution. We see larger differences in the gamma parameters, which represent participants' tendencies to obtain a boundary response on each cognitive test. These parameters have become more extreme in Study 2, with the lower parameter (representing a lower boundary response) becoming more negative and the upper parameter (representing an upper boundary response) becoming more positive. Because boundary responses are relatively rare, it is reasonable to assume that the larger sample sizes from Study 2 led to these differences. Specifically, larger sample size reduced the amount of shrinkage in the gamma parameters, leading the model to produce estimates that were more extreme.

\begin{figure}
  \caption{Study 2 posterior mean test information curve (black line) versus posterior distribution of Study 1 test information curves (light blue lines).} \label{fig:infdist}
\begin{knitrout}\footnotesize
\definecolor{shadecolor}{rgb}{0.969, 0.969, 0.969}\color{fgcolor}

{\centering \includegraphics[width=\maxwidth]{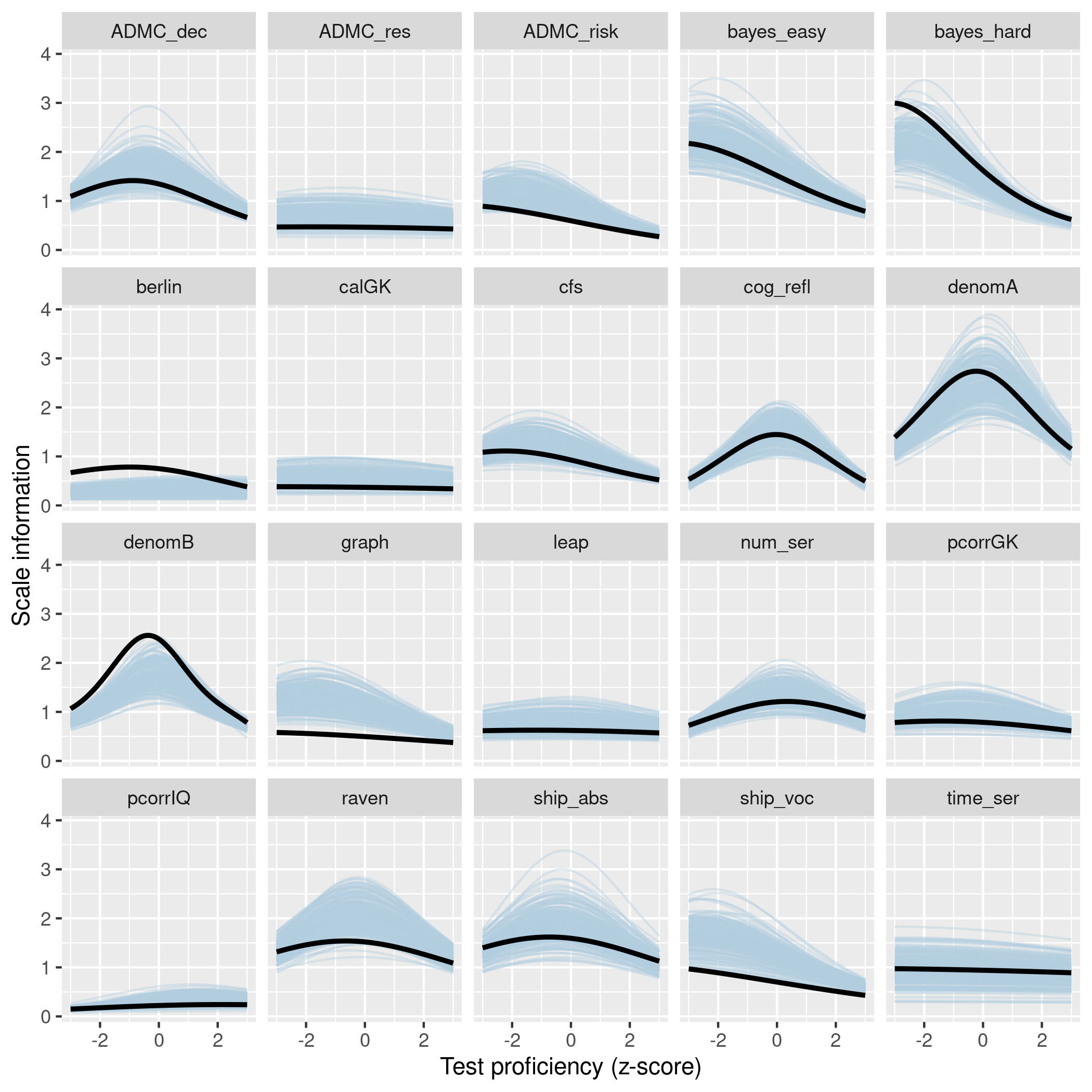} 

}

\end{knitrout}
\end{figure}

We next assess how these test parameter estimates combine to yield test information curves. Figure~\ref{fig:infdist} overlays the posterior mean test information curve from Study 2 (black line) on top of the posterior distribution from Study 1 (blue lines). Berlin Numeracy and Denominator Neglect (B) stand out as being slightly more informative in Study 2, while Graph Literacy is slightly less informative. But the figure shows that the Study 2 information curves generally lie in the Study 1 posterior distributions.

\begin{figure}
  \caption{Study 2 participant scores estimated directly within the Study 2 model (x-axis) versus scored using the Study 1 test estimates (y-axis).} \label{fig:perscomp}
\begin{knitrout}\footnotesize
\definecolor{shadecolor}{rgb}{0.969, 0.969, 0.969}\color{fgcolor}

{\centering \includegraphics[width=\maxwidth]{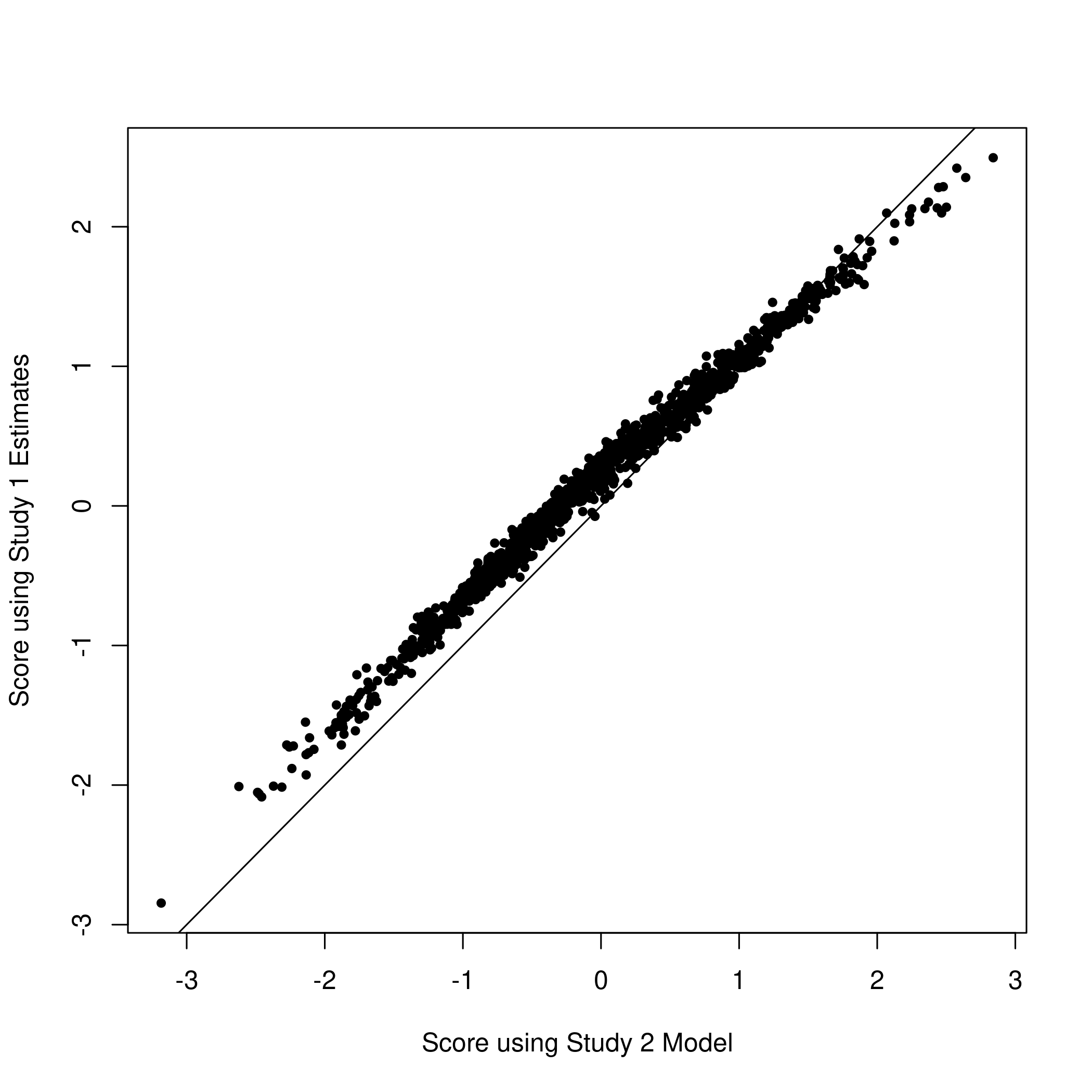} 

}

\end{knitrout}
\end{figure}

Next, we compare the estimated participant scores from the Study 2 model to the scores that involved Study 1 test parameter estimates. The comparison is shown in Figure~\ref{fig:perscomp}, where scores from the Study 2 model are on the x-axis, scores using the Study 1 estimates are on the y-axis, and each point is a participant. We see strong agreement in these scoring methods, with a correlation of~0.99. The scores using Study 1 parameter estimates are somewhat less extreme than the scores from the Study 2 model, reflecting a reluctance to assign exceptionally good or bad scores. We think this is again because the Study 1 estimates are based on a smaller sample size, and the gamma parameter estimates were less extreme than those of Study 2.

\begin{figure}
  \caption{Relationship between Study 2 cognitive test scores (x-axis) and Study 2 forecasting scores (y-axis).} \label{fig:st2prevsfore}
\begin{knitrout}\footnotesize
\definecolor{shadecolor}{rgb}{0.969, 0.969, 0.969}\color{fgcolor}

{\centering \includegraphics[width=\maxwidth]{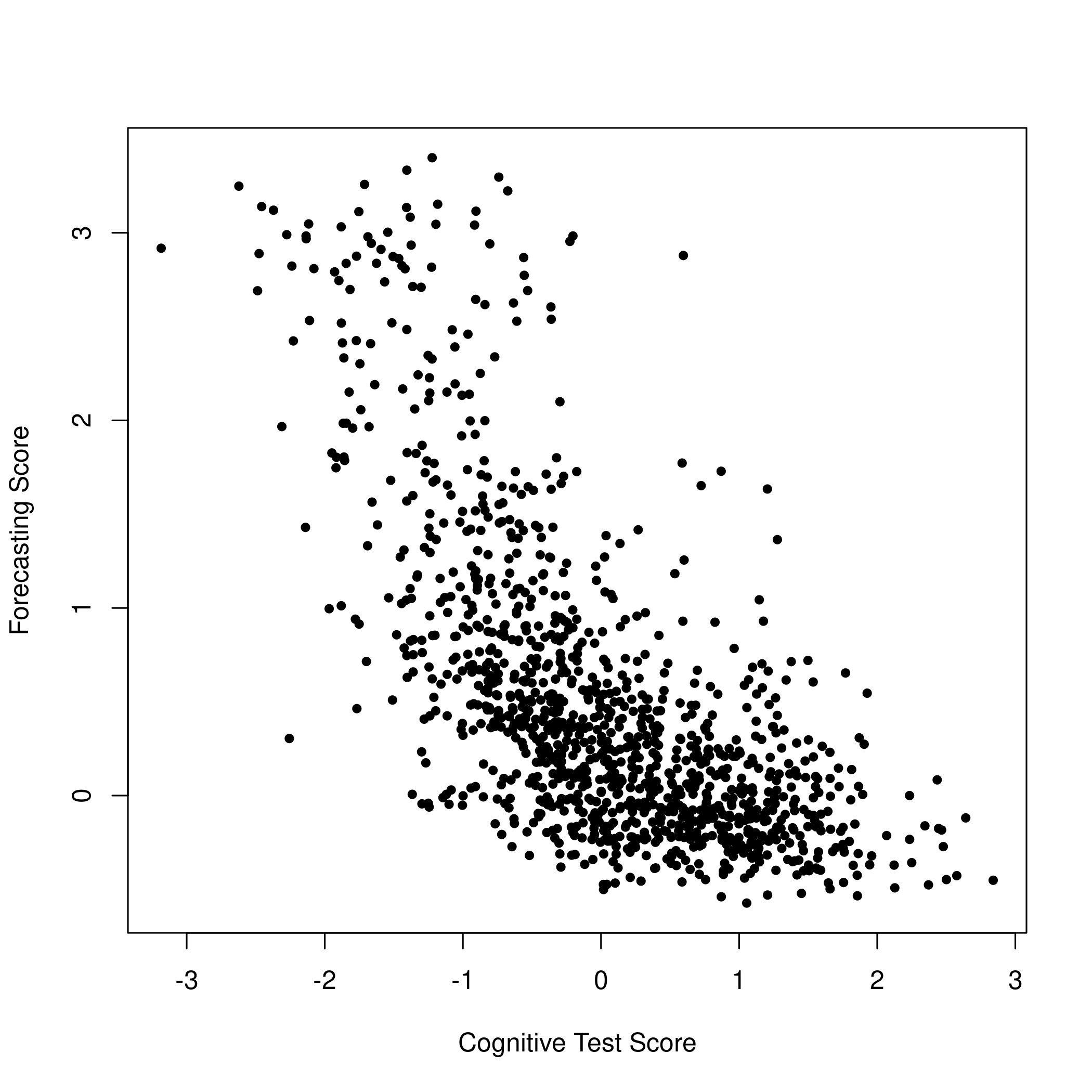} 

}

\end{knitrout}
\end{figure}

Regardless of which type of score we use, the correlation with the participants' later forecasting scores is near $-0.67$. This relationship is shown in Figure~\ref{fig:st2prevsfore}, where it should be recalled that higher cognitive test scores are better and lower forecasting scores are better. The relationship remains strong and similar to that of Study 1.

\subsection{Summary}
In fitting our model to the Study 2 data, we found that (i) question parameter estimates generally agreed with those of Study 1; (ii) information curves generally agreed with Study 1; (iii) participant scores were nearly the same, regardless of whether we directly estimated the scores in the Study 2 model or estimated the scores using Study 1 estimates; and (iv) participant scores remained strongly correlated with forecasting performance. The third result is especially important for adaptive testing because, if we can use previous estimates of question parameters, we can quickly score participants in real time and select cognitive tests accordingly. The fourth result is especially important for the overall goal of the study, which is to use cognitive tasks to predict forecasting performance. In the next section, we use results and data from the two studies to implement an adaptive testing scenario.

\section{Adaptive Cognitive Testing}
We now combine results from the previous sections to develop an adaptive testing procedure, and to apply the adaptive testing procedure to the Study 1 data. We examine how the procedure tailors the sequence of cognitive tests to each participant, and the time that it takes to achieve reliable estimates of a participant's performance.

\subsection{Method}
The adaptive testing procedure that we implemented includes the following steps:
\begin{enumerate}
\item Have all participants start with the test that is most informative for the average participant.
\item Score each participant using all tests that they have completed so far.
\item Based on each participant's running score, administer the test that is expected to be most informative and that has not yet been administered to the participant.
\item Repeat Steps 2 and 3 until all tests have been administered.
\end{enumerate}
In Step 2 above, we use the estimated question parameters from Study 2 to score each participant (which leads to empirical Bayes estimates of the $\theta_i$ parameters). In Step 3, we use two types of information from Study 2: the usual test information (shown in Figure~\ref{fig:infdist}), and test information per minute (similar to Figure~\ref{fig:postsampmin}). The latter may be useful because it attempts to gain the most information about each participant in the shortest amount of time.

We apply the adaptive testing procedure to all 146 Study 1 participants with complete data. We use 18 of the 20 test scores, excluding the Impossible Question and General Knowledge Calibration scores, both of which come from the Impossible Question task alongside the General Knowledge Percent Correct score. We excluded these scores because (i) they were generally uninformative and (ii) they lead to problems with our timing calculations (i.e., three scores arise simultaneously from one task).

At the start of the adaptive testing procedure, we pretend that the participants have not completed any cognitive tests. We then sequentially enter participants' scores in to the adaptive testing procedure, based on the tests that the procedure selects. In this way, we can examine how the procedure would have behaved, had we actually used it during the Study 1 data collection.

\subsection{Results}
\begin{figure}
  \caption{Sequences of tests administered in Study 1. Darker points and lines indicate that more people received a specific sequence.} \label{fig:adseq}
\begin{knitrout}\footnotesize
\definecolor{shadecolor}{rgb}{0.969, 0.969, 0.969}\color{fgcolor}

{\centering \includegraphics[width=\maxwidth]{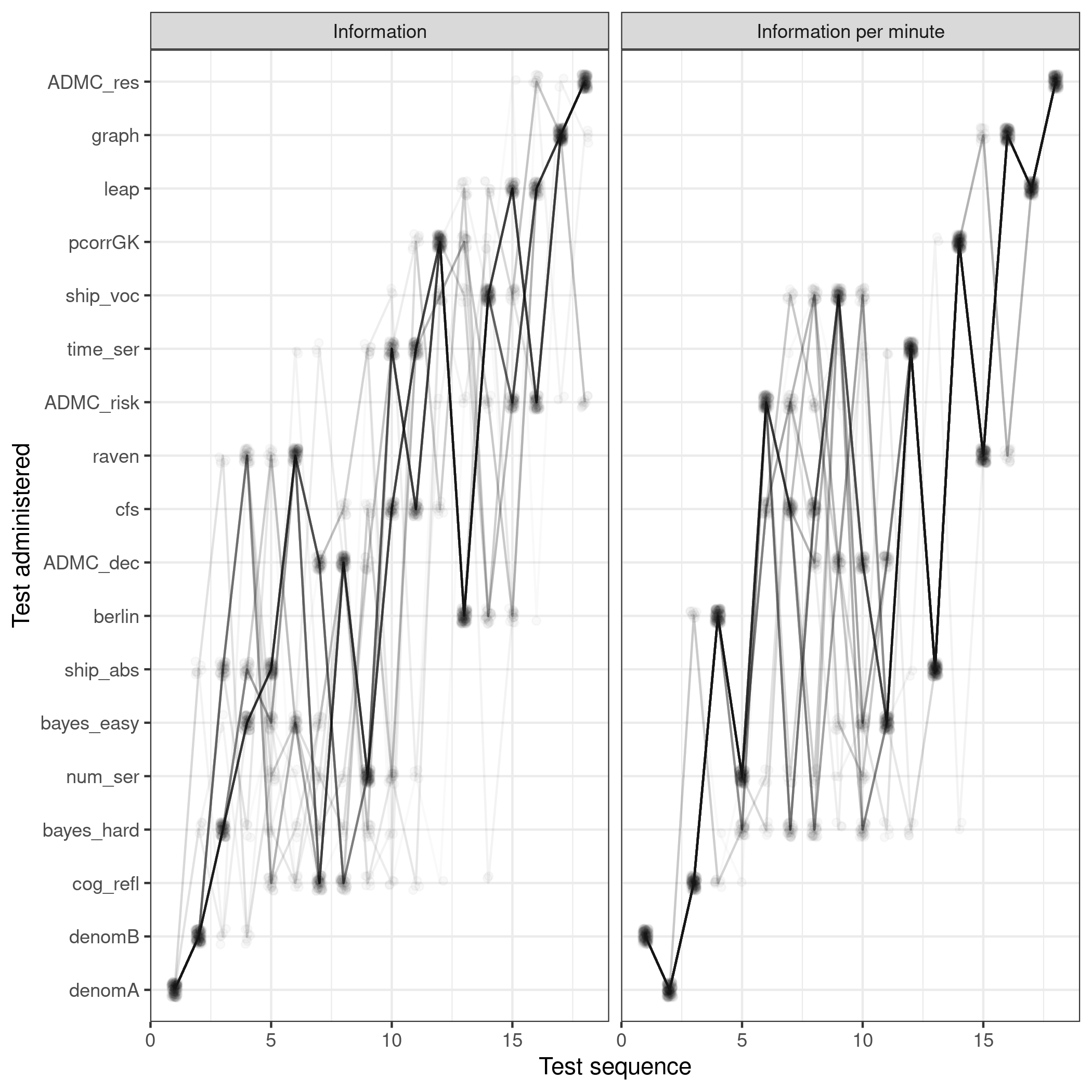} 

}

\end{knitrout}
\end{figure}

Figure~\ref{fig:adseq} shows the order in which cognitive tests were selected for each individual. The x-axis represents the order/time of administration, where 1 is the first test selected and 18 is the last test selected, the y-axis shows the tests themselves, and each point represents a specific cognitive test selected for a specific participant at a specific timepoint. The left panel is for the adaptive testing procedure that uses test information, and the right panel is for the analogous procedure that uses test information per minute.

We see many similarities across the two panels. Denominator Neglect (A or B) is selected first, because these are the two most informative tests. Denominator Neglect A takes a slightly longer time to complete than Denominator Neglect B, and so the order of administration flips across the two panels (A before B using test information; B before A using test information per minute). The cognitive tests administered last are also similar across the two panels: the ``Resistance to Framing'' part of the Adult Decision Making Competence scale, Graph Literacy, and Leapfrog tend to be administered last.

There are also some differences across the two panels, especially for cognitive tests that took a long time to complete. The ``Abstraction'' part of the Shipley General Intelligence test arises early in the left panel, but it becomes twelfth or thirteenth administered in the right panel. This is because this test takes nearly 10 minutes to complete, which is a long time compared to the other tests. Similarly, the Raven Matrices are highly informative but also take longer to complete (about 11 minutes), and so this task appears further right in the right panel, compared to other tests.

\begin{figure}
  \caption{Testing time (x-axis) versus correlation between cognitive test proficiency and forecasting accuracy (y-axis).} \label{fig:timevscor}
\begin{knitrout}\footnotesize
\definecolor{shadecolor}{rgb}{0.969, 0.969, 0.969}\color{fgcolor}

{\centering \includegraphics[width=\maxwidth]{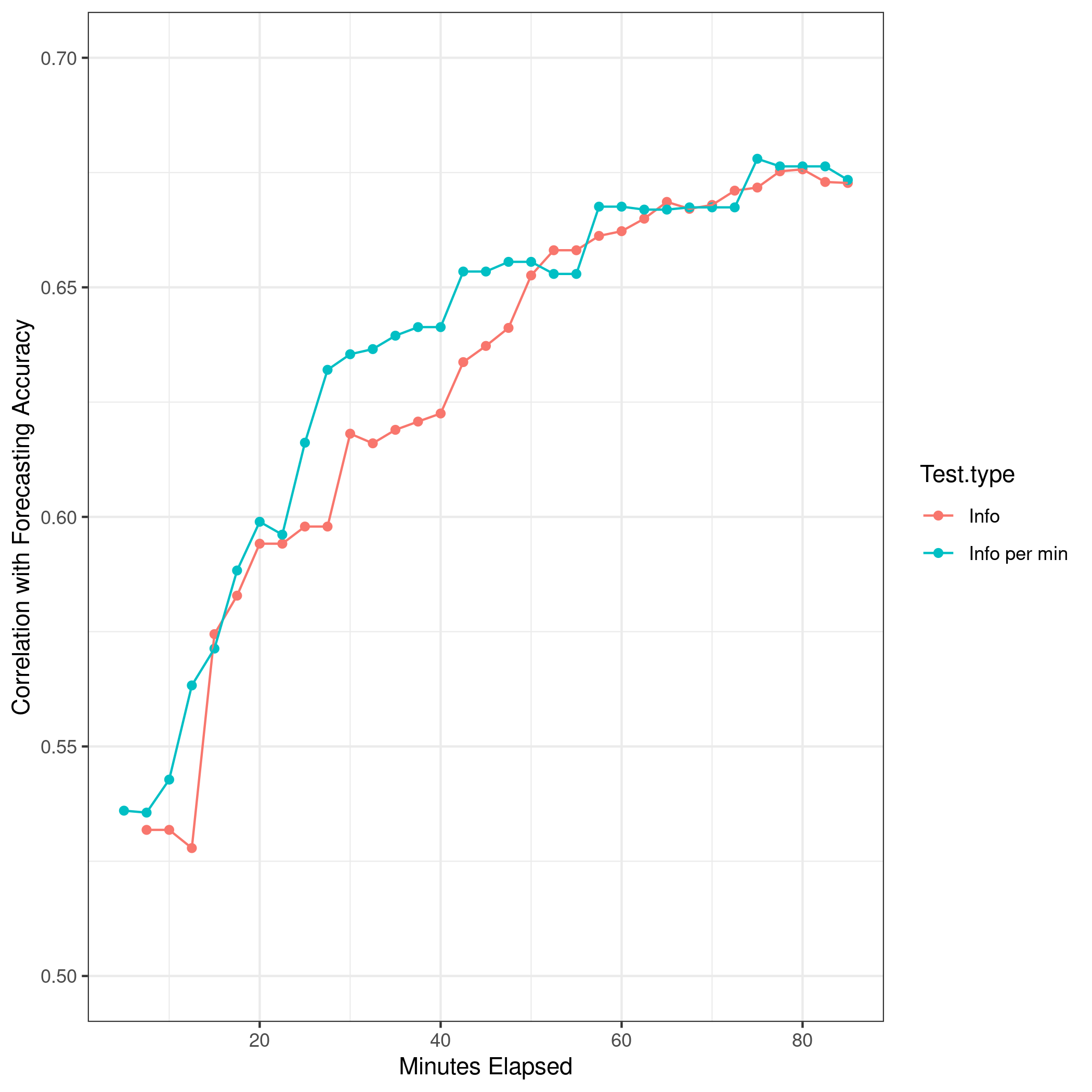} 

}

\end{knitrout}
\end{figure}

Figure~\ref{fig:timevscor} shows how the correlation between cognitive test proficiency and forecasting accuracy evolves over the course of each adaptive test. There are two lines, one for the adaptive test that uses test information alone, and another for the adaptive test that uses test information per minute. We see that the initial points on the left already have correlations above 0.5 (in absolute value); that is, the Denominator Neglect scores are already highly correlated with forecasting accuracy. As we continue along the x-axis, the ``information per minute'' line has an advantage from about 25 minutes elapsed to 50 minutes elapsed, illustrating an advantage of prioritizing tests that can be quickly administered. We also see that the lines flatten out near 60 minutes, indicating that we can safely eliminate the final 30 minutes of cognitive testing. The adaptive testing procedure provides a principled way to decide on the amount of time that should be devoted to cognitive testing. Namely, researchers can consider whether 30 extra minutes of cognitive testing is worthwhile, if those 30 extra minutes increase the correlation with forecasting accuracy from 0.60 to 0.67.

\begin{figure}
  \caption{Recommended test selections, for participants of different proficiencies (x-axis) and for different cumulative lengths of testing (y-axis).} \label{fig:timerec}
\begin{knitrout}\footnotesize
\definecolor{shadecolor}{rgb}{0.969, 0.969, 0.969}\color{fgcolor}

{\centering \includegraphics[width=\maxwidth]{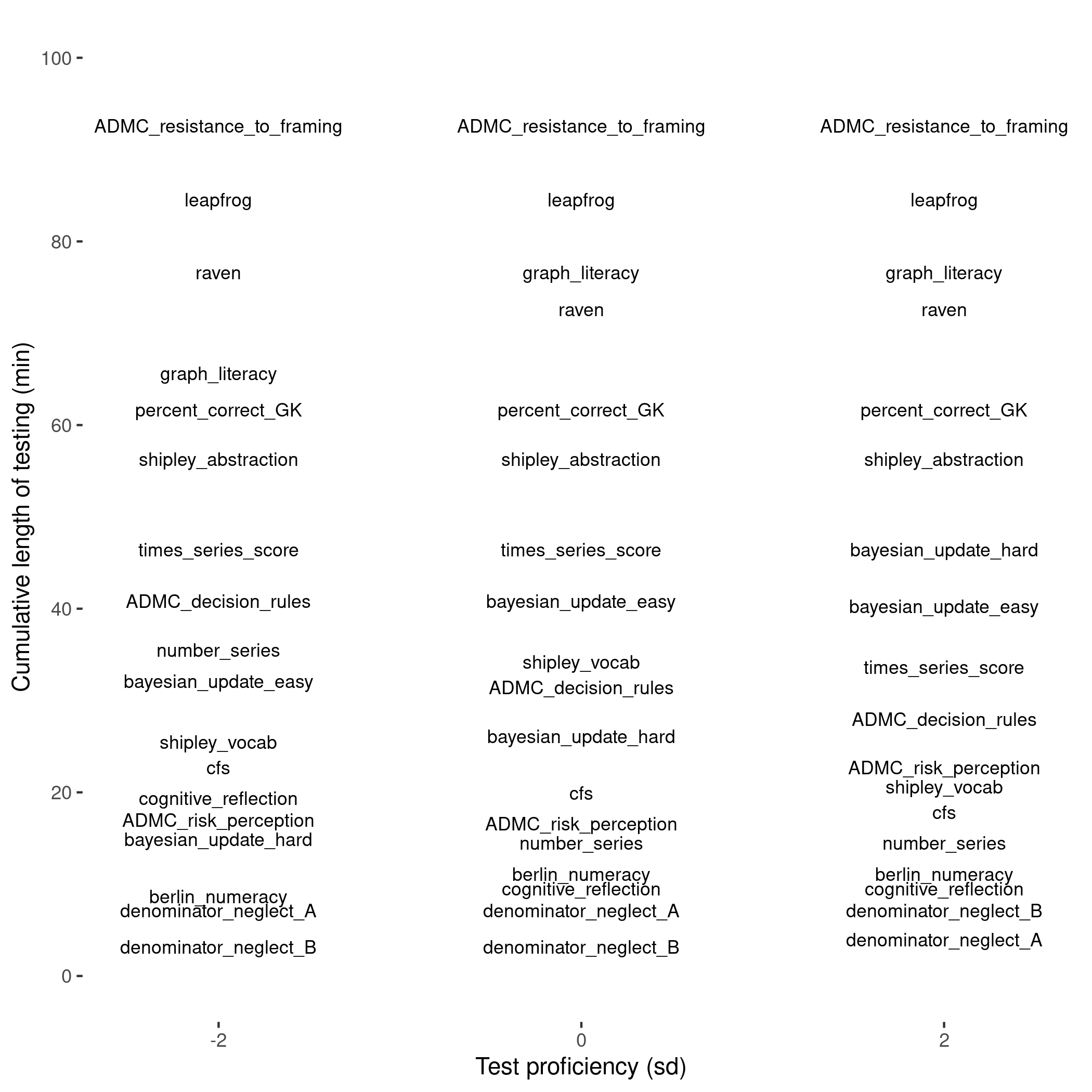} 

}

\end{knitrout}
\end{figure}

Because Figure~\ref{fig:adseq} showed that tests tend to be administered in the same order, we may elect to prescribe a single ordering of tests for all participants, as opposed to tailoring an order to each participant. Alternatively, we may prescribe three separate orderings for low-, average-, and high-proficiency participants, respectively. Figure~\ref{fig:timerec} shows such prescriptions, where cognitive test proficiency is on the x-axis and testing time is on the y-axis. These prescriptions are based on the ``information per minute'' metric, and therefore the longer tasks usually appear near the end of the ordering. Using the figure, we can draw a horizontal line at a desired time, and then administer all tests below that line. We see that, if we have 10 minutes or less, we would generally administer the Denominator Neglect and Cognitive Reflection tests, whereas if we have 20 minutes, we would administer 6 or 7 tests. The prescriptions on the right side are likely to be of primary interest, because these are the tests that are most informative about good forecasters.

\section{General Discussion}
In this paper, we applied item response models and adaptive testing principles to yield insights into the relationship between cognitive tests and forecasting proficiency. We applied a beta item response model to cognitive test scores and estimated the test information curves, which helped us discern the suitability of each test to participants of varying proficiencies. We used the test information curves to select informative tests and discard uninformative tests, reducing the time required to administer the tests without compromising predictive accuracy. We verified that the selected tests generalized to a second, independent dataset, and we then showed how the test information curves could be used to develop an adaptive testing procedure that works in real time. In this section, we consider limitations and future developments.

\paragraph{Limitations}
While our model performance was reasonable, there are places where it could be improved and where our conclusions should be qualified. We expand on these points below.

First, the beta item response model required us to set bounds on each cognitive test's possible scores, and the exact bounds are not always well defined. For example, the Leapfrog task has theoretical boundaries of 0 and 400, but it is nearly impossible to achieve a score near either bound. We therefore used empirical boundaries based on the Study 1 data. This use of empirical boundaries limits the possible scores that could be achieved in future testing. For example, if a Study 2 participant scored higher than the empirical Study 1 upper bound, then that participant would be treated as though their score equaled the empirical Study 1 upper bound. This potentially limits our ability to distinguish between exceptional participants (exceptionally proficient or improficient). An alternative would be to scrap the boundaries by employing a Gaussian IRT model in place of the beta item response model. The information functions under the Gaussian model are flat, though, which prevents us from tailoring tests to different participants.

Related to the boundary issue, some of the cognitive tests are scored in terms of the number of questions answered correctly. For these tests, it may be better to use a binomial item response model (setting the binomial $n$ parameter to the total number of questions) instead of a beta item response model. Use of the binomial model would remove the need to set bounds for some cognitive tests. The information functions would then be of a different form, depending on whether the cognitive test uses a beta or a binomial likelihood. These different information functions may further heighten our ability to predict forecasting accuracy.

Finally, as we mentioned in the Introduction, our beta item response model assumed a single dimension of proficiency across all the cognitive tests. This assumption helps limit the complexity of the model, which potentially improves model predictions (including test information) for new datasets \cite<e.g.,>{mersaw20}. But this assumption is sure to be incorrect, which qualifies our conclusions about the test information functions: there remains the possibility that some of the low-information tests from our study are informative about certain aspects of forecasting, but we cannot see that when we lump together all the cognitive tests in one dimension. It would be possible to fit beta item response models with multiple dimensions of proficiency, with each test having a unique information function for each dimension. In designing an adaptive testing procedure, we then must decide which dimension(s) of proficiency to prioritize. There are many possibilities here, both in terms of the number of proficiency dimensions and in terms of cognitive test selection. The simplicity of a unidimensional model is advantageous, in that it reduces the forking paths of modeling attributes and adaptive test designs.

\paragraph{Further Applications}
Along with multidimensional item response models, future work could explore applying these models to related datasets. First, while each participant currently receives a single score on each cognitive test, we could consider applying an item response model to the individual questions in each cognitive test. If there are a few highly informative questions in each test, such an item response model would help us isolate those questions, leading to an adaptive test that takes even less time than those examined here. Second, we may apply the adaptive testing approach to ``repeatable'' forecasting questions, such as Time Series questions with differing time horizons. We could estimate the information of each forecasting question, and then present different questions to forecasters based on their forecasting performance to date. Combining this idea with multidimensional item response models, we may select questions for each forecaster depending on the specific dimensions along which each forecaster excels. We anticipate that large datasets will be necessary to estimate the parameters of such a model with enough precision to be useful.

\paragraph{Summary}
The models and procedures used in this paper appear fruitful for continued application to cognitive tests, to forecasting questions, and to combinations thereof. They help us find the best tests for assessing forecasters' proficiency, and they allow us to develop shorter adaptive tests that are as good as longer tests. The procedures are also efficient enough to be used in real time, after a model has been fit to a base dataset. We look forward to further developments along these lines in the future.

\section*{Computational Details}

All results were obtained using the \proglang{R}~system for statistical computing \cite{rprog},
version~4.5.0, especially relying on the {\em rstan} \cite{rstan}, {\em bayesplot} \cite{bayesplot}, {\em ggplot2} \cite{ggplot2}, and {\em targets} \cite{targets} packages.

\appendix

\section{Additional Results}
\begin{figure}
  \caption{Correlations of cognitive test scores.} \label{fig:corhist}
\begin{knitrout}\footnotesize
\definecolor{shadecolor}{rgb}{0.969, 0.969, 0.969}\color{fgcolor}

{\centering \includegraphics[width=\maxwidth]{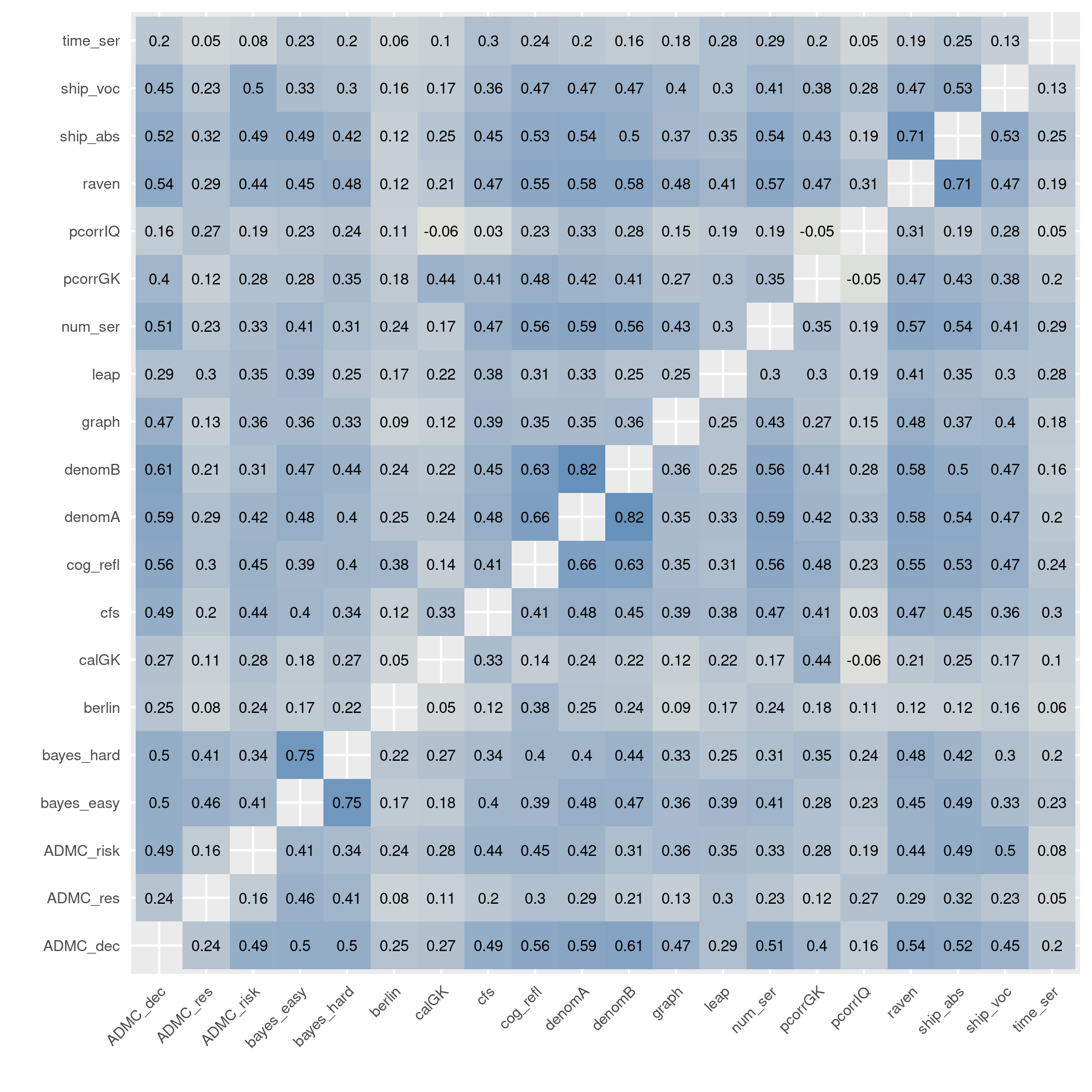} 

}

\end{knitrout}
\end{figure}

Figure~\ref{fig:corhist} shows pairwise correlations of all the test scores from Study 1. Darker blue squares depict stronger correlations, and lighter squares depict correlations near 0.

\end{document}